# Tracking-free Determination of Microparticle Motion from Image Variance

Harish Sasikumar[1], and Manoj M. Varma[1,2,*]

[1]Center for Nano Science and Engineering, Indian Institute of Science, Bangalore, 560012, India

[2]Robert Bosch Center for Cyber Physical Systems, Indian Institute of Science, Bangalore, 560012, India

**Abstract**

In this work, we use the standard deviation of image pixel intensity to analyse the speed, direction and surface-interaction of microparticles in fluid. First, we present an analytical model for estimating the total variance in the image space for directed or diffusive motion of microparticles and show that this measure is correlated to the density and speed of the particles. The analytical model was found to have good agreement with numerical simulations for low particle density. Then, using only the local image variance we obtain the magnitude and direction of the particle velocity in a rectangular microfluidic channel, closely matching the theoretical profile. Further, we also demonstrate the application of this method as a probe for particle-surface interactions by extracting the differences in distribution and time-evolution of image variance from mobile microparticles adhering to different surfaces. We believe that the image variance based method described here presents an addition to the suite of tracking-free techniques such as Differential Dynamic Microscopy (DDM) to extract motility parameters from video data.

**Introduction**

Ability to quantify motion at microscale is of fundamental importance in understanding several phenomena involving Brownian and phoretic motion of particles as well as their interactions with materials they come in contact with. A straight-forward method for these studies is particle-tracking, where the positions of individual particles are tracked over time followed by computational analysis of particle trajectories, for instance, as in micro particle velocimetry [1]. These techniques generally involve the use of fluorescent or other tracer particles and computational extraction of particle position as a function of time. Often elaborate imaging techniques and autocorrelation of the images are used to estimate the particle velocities [2]. Though information of particle dynamics is present in its entirety in these methods, there are two aspects which motivates development of tracking-free methods to characterize micro-scale motion. Firstly, tracking data is voluminous and often one is only interested in summary statistics - central tendencies, measures of dispersions etc. of particle locations, which represent a significant compression of information contained in the raw particle trajectory data. Secondly, tracking is computationally intensive and prone to errors and is particularly challenging in crowded environments. Therefore, obtaining summary statistics of motion by tracking-free methods is desirable.

Among tracking-free methods, Differential Dynamic Microscopy (DDM) has emerged as a popular technique for analysing the dynamics of particles – both inanimate [3, 4] and animate [5, 6]. Apart from saving time and effort, this method avoids the errors associated with particle tracking and provides a reliable method to estimate the average dynamics of microscopic particles [7]. DDM studies can work well even at the optical resolution limits and reveal the finer workings of many microscopic systems. For instance, usefulness of DDM in differentiating directed motion and random diffusion of sub-diffraction particles in crowded and noisy environment has made it a suitable procedure for studying intracellular transport mechanism [8]. In conventional DDM, the analysis is often done in k-space by assuming the isotropy of the image (or differential image). This allows one to take azimuthal averages in the Fourier domain to isolate the different Fourier components. There are only a few deviations to this general approach which are reported; for instance, Differential Variance Analysis (DVA) [9] has been used to monitor dynamic spatial heterogeneities in amorphous materials. There are also methods, such as the one reported by Mathias et al. [10], to quantify the alignment of anisotropic colloids to external fields by extracting the directionality using DDM. In this case, a more generalized form of DDM is derived, avoiding the azimuthal averaging, maintaining the vector nature of the wave vector. It is to be noted that even here, it is the global orientational order parameter of the system which is extracted.

---

[*] Author to whom correspondence should be addressed.  Electronic mail: mvarma@iisc.ac.in

Anisotropy in the differential image data, arising from spatial variations in the speed or direction of motion is common and is important to be captured. Micro particle image velocimetry (mPIV) techniques have been applied to get the speed and direction of micro-scale flows [11]. This technique requires the autocorrelation of image fields to estimate the spatial shifts and can be computationally expensive [12]. In this paper, we show that the variance (or standard deviation) of the pixel intensity, which can be easily extracted from video data, can be used to obtain velocity fields. This approach does not require calculations of correlations or transforming data to fourier domain. By dividing the video data spatially and temporally, one can obtain the speed and direction of micro-particles in the system. The inherent heterogeneities of the system are also captured due to this spatial compartmentalization. In addition to motility, one can also use pixel variance to gather information about the nature of interaction of particle with surfaces in contact with it. Studying these interactions is useful in developing advanced surfaces like polyelectrolyte (PE) brushes, where often techniques like optical tweezers and AFM colloidal probes are used in this regard [13]. These surface interactions also affect the Brownian motion of micro particles in contact with the surfaces resulting in variations in the number of mobile particles, their velocity distributions as well as the time evolution of their velocities. Hence, the statistics of the movements of microparticles can in turn be used in studying the properties of surfaces. Recently such fluctuations were used for sensing the position and orientation of single, bound viruses [14].

The paper is organized in the following manner. In the first section, we present theoretical and numerical studies to clarify the process of using pixel variance to estimate motility parameters for directed as well as diffusive motion. We then show an example of using pixel variance to obtain the velocity profile from video recording of flow in microchannels with rectangular cross-section. Finally, we show how temporal evolution of the pixel variance informs one about the nature of interaction between a microparticle and the surface it contacts.

## Theory and Simulation

### Theoretical Model

Consider an imaging plane of area $a \times b$ pixels with p particles at random locations. Each pixel in the area has a value $x_{i,j}$, addressed by its location $(i, j)$, which is an element in the set $G = \{(1,1), (1,2), \ldots (1, b), (2,1), \ldots, (a, b)\}$ with a cardinality ab.

For simplicity, we assume monochrome images with a dimension scaling factor of $\gamma$ (usually measured in the units $\frac{\mu m}{pixel}$), acquired at a frame rate of $f$ (usually measured in the units $\frac{frames}{second}$). The acquisition is done for a total time of T seconds, which corresponds to $N (= Tf)$ frames. In the analytical model, we assume that the particles are sparsely distributed so that during the observation time, their paths do not collide or cross over. Further, each particle is assumed to be occupying only one pixel in the observation area. (These two assumptions are removed in the simulation model). The particle is considered to be opaque which makes the occupied pixel to have a lower value (l), than that of an unoccupied pixel (h).

For obtaining an analytical model for the linear motion, we assume a particle velocity of v which creates a scaled particle velocity $v_s$ in the images, given by

$$v_s = \frac{v}{\gamma f} = \frac{1}{n} \tag{1}$$

where n is the number of time frames for which the particle appears stationary before moving on to the next pixel along its direction of motion. It is to be noted that $v_s$ has unit $\frac{pixels}{frames}$. We assume that the total number of frames captured is an integral multiple of n, i.e $N = mn$, where m is a positive integer. In effect, $m$ is the total displacement of the particle (in unit of pixels) during the total observation time. When observed for N time frames, the pixels along the trajectories of particles change, whereas the value of the remaining pixels remain constant (h). Hence, the average value of a pixels in the imaging plane is given by,

$$\overline{(x_{i,j})} = \begin{cases} x_{path} = h - \frac{\Delta}{Nv_{sc}} & ; for\ x_{i,j} \in S_{i,j} \\ x_{blank} = h & ; for\ x_{i,j} \notin S_{i,j} \end{cases} \tag{2}$$

Where $S_{i,j}$ is the set of pixels where the particle has occupied during the observation and $\Delta$ is the difference in pixel values, given as $h - l$. The above expression indicates a velocity dependent histogram of the averages of pixels. Similarly, the standard deviation of the pixels in the imaging plane also has a velocity dependent histogram given by,

$$\sigma(x_{i,j}) = \begin{cases} \frac{|\Delta|}{\sqrt{Nv_s}} & for\ x_{i,j} \in S_{i,j} \\ 0 & for\ x_{i,j} \notin S_{i,j} \end{cases} \tag{3}$$

Summing up the standard deviations of all the pixels in the imaging plane, we can define and calculate a global parameter S given by,

$$S = \sum_{(i,j) \in G} \sigma(x_{i,j}) = p|\Delta|\sqrt{Nv_s} \tag{4}$$

Note that, here p is the number of particles. Though the equations developed so far provide an insight into the correlation between pixel standard deviations and velocity of the particles, they can be insufficient for predicting experimental outcomes. It is also worth noting that for particles with uniform velocity, S is independent of the area of simulation and is linearly dependent on the number of particles. However, in real experiments, these equations are inaccurate because the particles often occupy multiple pixels and their trajectories can intersect. Hence, a simulation model was developed which takes care of longer simulation times with trajectory crossings, and multi-pixel particles.

For modelling systems with circularly symmetric larger (multi-pixel) particles, the images have to be considered as airy functions as they are the point spread function of a diffraction-limited particle. However, for mathematical convenience, we use the Gaussian distribution to approximate this function [15].

$$x_{i,j}^{(G)} = Ae^{-\frac{(i-i_r)^2+(j-j_r)^2}{2R^2}} \tag{5}$$

Where $x_{i,j}$ is the value of the pixel at $(i, j)$ for the Gaussian distribution with center at $(i_r, j_r)$ and the variance R. See Section 2 of supplementary information (SI) for the comparison of multi-pixel simulation model and experimentally observed micro particle. For the case where the total displacement of the multipixel particle is larger than the particle dimension ($Nv_s \gg R$), the average pixel value ($\overline{x_{i,j}^{(G)}}$) is given by

$$\overline{x_{i,j}^{(G)}} = A\sqrt{\pi}\frac{R}{N}e^{-\frac{i^2}{4R^2}} \tag{6}$$

Similarly, the mean square displacement $\overline{\left(x_{i,j}^{G}\right)^2}$ can be calculated, using which the variance of individual pixels ($\sigma^2(x_{i,j})$) can be derived as,

$$\sigma^2(x_{i,j}) = A^2\left(\sqrt{\frac{\pi}{2}}\frac{R}{N} - \pi\frac{R^2}{N^2}\right)e^{-\frac{i^2}{4R^2}} \tag{7}$$

And the sum of the standard deviations of all pixels due to p particles are given by

$$S^{(G)} = p\sigma = \frac{Apv}{\gamma f}\sqrt{\sqrt{\frac{\pi^3}{2}}NR^3\left(1 - \sqrt{2\pi}\frac{R}{N}\right)} \tag{8}$$

Please refer to supplementary information (SI) section 3.1 for detailed derivation of Eq. 8.

**Simulation Model**

The simulation parameters were chosen to closely match the experimental conditions, namely, the size of the observation area and resolution of the imaging camera. For representative purposes, FIG. 1 (a) to (d) shows a simulation area with a dimension of 85 pixel × 85 pixel. For modelling uniformly illuminated background with opaque microparticles, a high value ($h = 1$) background was populated by pixels occupied by microparticles at lower values. To verify the theoretical approximations, initially the particles were simulated as single pixel model, given by

$$x_{i,j} = \begin{cases} 1 \text{ ; for } x_{i,j} \notin \{p_{i,j}\} \\ 0 \text{ ; for } x_{i,j} \in \{p_{i,j}\} \end{cases} \tag{9}$$

where the $p_{i,j}$ is the set of central locations of the particles. FIG. 1(a) shows an area of simulation of 85 pixel × 85 pixel with 10 randomly distributed particles represented by single pixel model. (b) shows an area of simulation of 85 pixel × 85 pixel with 10 particles represented by multipixel model with $R = 1$. Periodic boundary conditions were applied to all the four boundaries of the simulation area, so that the number of particles is conserved despite the free movement of particles. After a fixed amount of time (represented by a simulation time of N time frames), the variance of each pixel is calculated as

$$\sigma^2(x_{i,j}) = \frac{1}{N} \sum_{n=1}^{N} \left(x_{i,j}(n) - \overline{x_{i,j}}\right)^2 \tag{10}$$

where $x_{i,j}(n)$ is the value of the pixel at the frame n and $\overline{x_{i,j}}$ is the time average value of the pixel given by

$$\overline{x_{i,j}} = \frac{1}{N} \sum_{n=1}^{N} x_{i,j}(n) \tag{11}$$

FIG. 1 (c) and (d) are the respective variances plots for single pixel and mutli-pixel models where the particles are simulated to move downwards by 60 contiguous positions from the locations in FIG. 1 (a) and (b). The periodic boundary condition is evident from the variance plots, where the particle trajectories ending at the bottom of the simulation area is restarted from the top. It is also apparent from these plots that the standard deviation plots align along the direction of the particle movement; vertical in this case. A global indicator of the particle movement, S, defined in equation 4 as the pixel standard deviation, was calculated. The simulations were done on a rectangular area of 1200 pixels high and 1600 pixels wide, a typical value closer to the image dimensions obtained from our camera. FIG. 1 (e) shows the comparison of the evolution of total standard deviation as predicted by the theory, the single-pixel (SP) model and the multi-pixel (MP) model. It can be seen that the theoretical predictions closely match the simulations at lower particle densities and shorter simulation times. At higher particle densities or when the simulation is done for a longer time, particle trajectories invariably cross and the total standard deviation predicted from the theoretical model overestimates the pixel variance as it assumes that the particle trajectories are mutually exclusive. With increasing particle number, the chance of trajectory crossings increases, resulting in larger overestimation of the pixel variance as seen in FIG. 1 (e). Similarly, the multi-pixel situation (i.e. when a particle covers multiple pixels in the image) exacerbates the effect of trajectory crossings and results in larger deviation between theory and simulation.

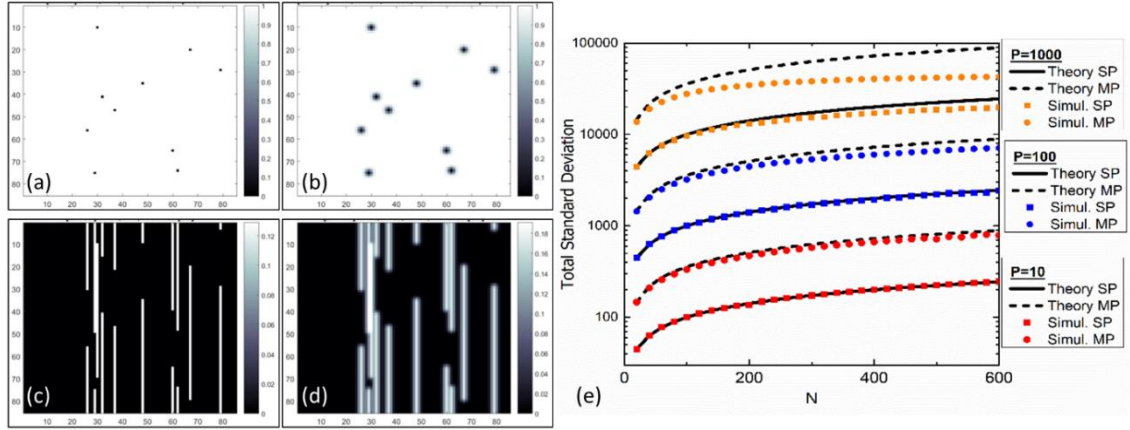

FIG. 1. (a) and (b) are representative simulation area of 85 pixel × 85 pixel using respectively, single pixel (SP) and Multi pixel (MP) models for 10 randomly placed particles. (c) and (d) are the standard deviation plots after the particles have moved 60 pixels downwards. (e) The evolution of total standard deviation, S as the simulation time, N increases for (P=) 10, 100 and 1000 number of particles in the 1200 pixel × 1600 pixel simulation area. R is taken as 1.5.

In case of two-dimensional Brownian movement, a two-dimensional random walk scenario was simulated i.e. at every time-step in the simulation, the subsequent positions of each of the particles were decided by random experiments of 4 equiprobable outcomes. Corresponding to each of these four outcomes, the particles were made to move in one of the four directions- up, down, left or right. FIG. 2 (a), (b), (c) and (d) are the standard deviation plot of such a simulation with initial configuration as in FIG. 1 (b) and for simulation times, N = 10, 100, 300 and 600, respectively.

FIG. 2 (e) is the evolution of total standard deviation, (similar to FIG. 1 (e)) for a system of 10, 100 and 1000 particles undergoing Brownian motion. For this, the position of the particle is updated at each time frame by translating it to one of the adjacent pixels, randomly. It can be seen that, for both single pixel and Gaussian models, S for Brownian motion stays lower than those in directed motion. In addition, even the slopes of the curves are lower than that of directed motion, making it apparent that the increase in pixel standard deviation is slower for systems with particles undergoing Brownian motion.

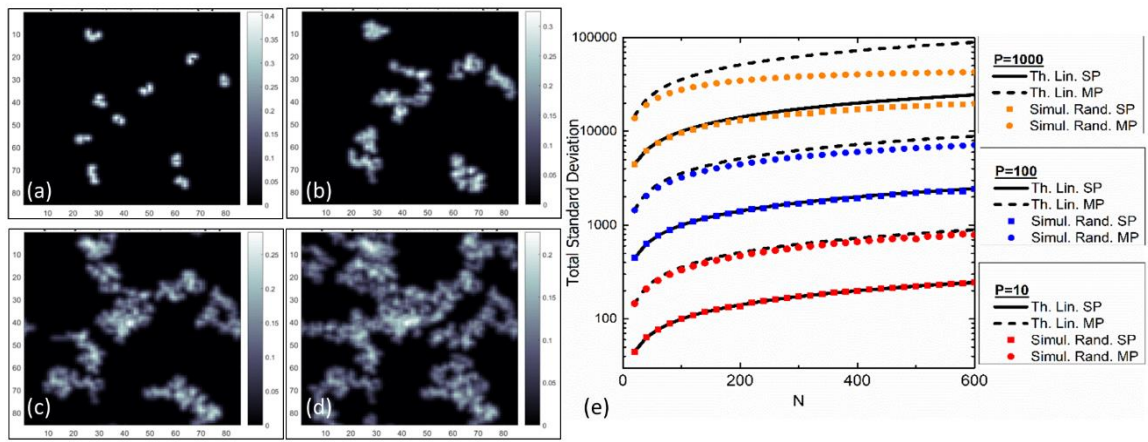

FIG. 2. Particles represented by Multipixel (MP) model undergoing Brownian motion for total time frames of N= (a) 10, (b) 100, (c) 200 and (d) 600 frames. (e) Time evolution of total standard deviation (S) for particles (derived theoretically (Th) and using simulation (Simul)) undergoing Brownian motion compared with the linear motion (Lin) of single pixel (SP) and multipixel (MP) particles. The particles are assumed to be moving by 1 pixel in every time frame.

# Experiments, Observations and Results

The movement of microparticles in a straight, laminar flow can be assumed as a good model system with directed motion. Hence, in order to verify the theoretical model, velocity profiles and flow directions were created using microfluidic devices in Polydimethylsiloxane (PDMS)[16]. The devices were fabricated using standard replica molding[17], with masters created using SU8 on Silicon (Refer Section 4 in SI text). A constant flow was maintained in the devices using a syringe pump. The device was observed using an optical microscope (Olympus, BX51M) with a 20x objective. The devices were imaged at 10 frames per second using a CCD monochrome camera (XM10) with an exposure time of 10 ms. The microparticles used in the experiments were polystyrene beads with diameter of 1 micron dispersed in an aqueous solution. These were procured from Sigma-Aldrich and used at a dilution corresponding to a number density of about $10^6$ mm$^{-3}$. Further details of the fabrication process, experimental setup and raw materials are given in SI section 4.

## Characterizing velocity profile of the particles

As shown in the schematic in FIG. 3 (a), a rectangular microfluidic channel (h = w = 220 μm) was created. The entire PDMS device was attached on a glass slide coated with gold. The pressure difference between either ends of the channel create a flow in the positive z direction with a flow profile varying along x and y. FIG. 3 (b) shows a 1200 pixel high and 1600 pixel wide microscopic image of this device. Videos of the flow was acquired (supplementary information as video1) which was later analyzed using ImageJ [18], an open-source software.

The standard deviations of the pixels for a time period of 1.5 seconds ($N$=15 frames) were calculated (Refer FIG. 3 (c)) for an X-Z plane close to the bottom surface. The total standard deviation (summed along z direction) at each of the x position was calculated using the formula

$$S(j) = \sum_{i=1}^{1200} \sigma(x_{i,j}) \qquad (12)$$

The velocity of the particles in each of the x position is then calculated using equation 8.

$$v(j) = \frac{\gamma f S(j)}{A p C}; \text{ where } C = \sqrt{\frac{\pi^3}{2} N R^3 \left(1 - \sqrt{2\pi}\frac{R}{N}\right)} \qquad (13)$$

$\gamma$ and $f$, being the dimension scaling factor and frame rate, are the properties of the imaging system. They were found to be 1.39 $\mu m/pixel$ and 10 frames/second, respectively. $A$ and $R$ are the height and standard deviation of the 2D Gaussian produced by the particle on the imaging system. They were calculated from the averages of 5 particles and were found to be 33.58 and 3.5 respectively. As $S(j)$ is summed along the $i$ direction, $p$ is the total number of particles (summed along $i$ directions) in each of the j positions. The estimate of this was found by dividing the number of particles with the width (in pixels) of the channel. The value of $p$ was found to be 1.9. $N$ is the number of time frames used in the calculations and is 15. The complete set of parameters used for computation are tabulated in Section 3.2 of SI text. FIG. 3 (d) shows the z-averaged velocity profile of the micro channel. As expected, a faster flow was observed along the centre of the microfluidic channels than the outer regions. In order to check the validity of the variance based determination of speeds, a limited number of particles were tracked and their speeds were determined from the tracked trajectories. These are shown as the red data points in FIG. 3(d). The close match between the magnitude of speed obtained using the variance method and single particle tracking reveals the validity of our method in extracting accurate motion parameters from recorded data. Further, we also compared the analytical solution to the Navier-Stokes equation for a pressure-driven steady state flow in the rectangular channel considered [19]. Velocity profile along z direction at regions close to the bottom surface (y = h/10) is given by

$$v_z(x) = \sum_{n,odd}^{\infty} \frac{1}{n^3} \left[ 1 - \frac{\cosh n\pi \frac{x}{h}}{\cosh n\pi \frac{w}{2h}} \right] \sin \frac{n\pi}{10} \qquad (14)$$

The first 100 non-zero terms in the above expression was used to plot the normalized Fourier sum approximation in FIG. 3 (d) and shows a qualitatively similar profile as obtained experimentally.

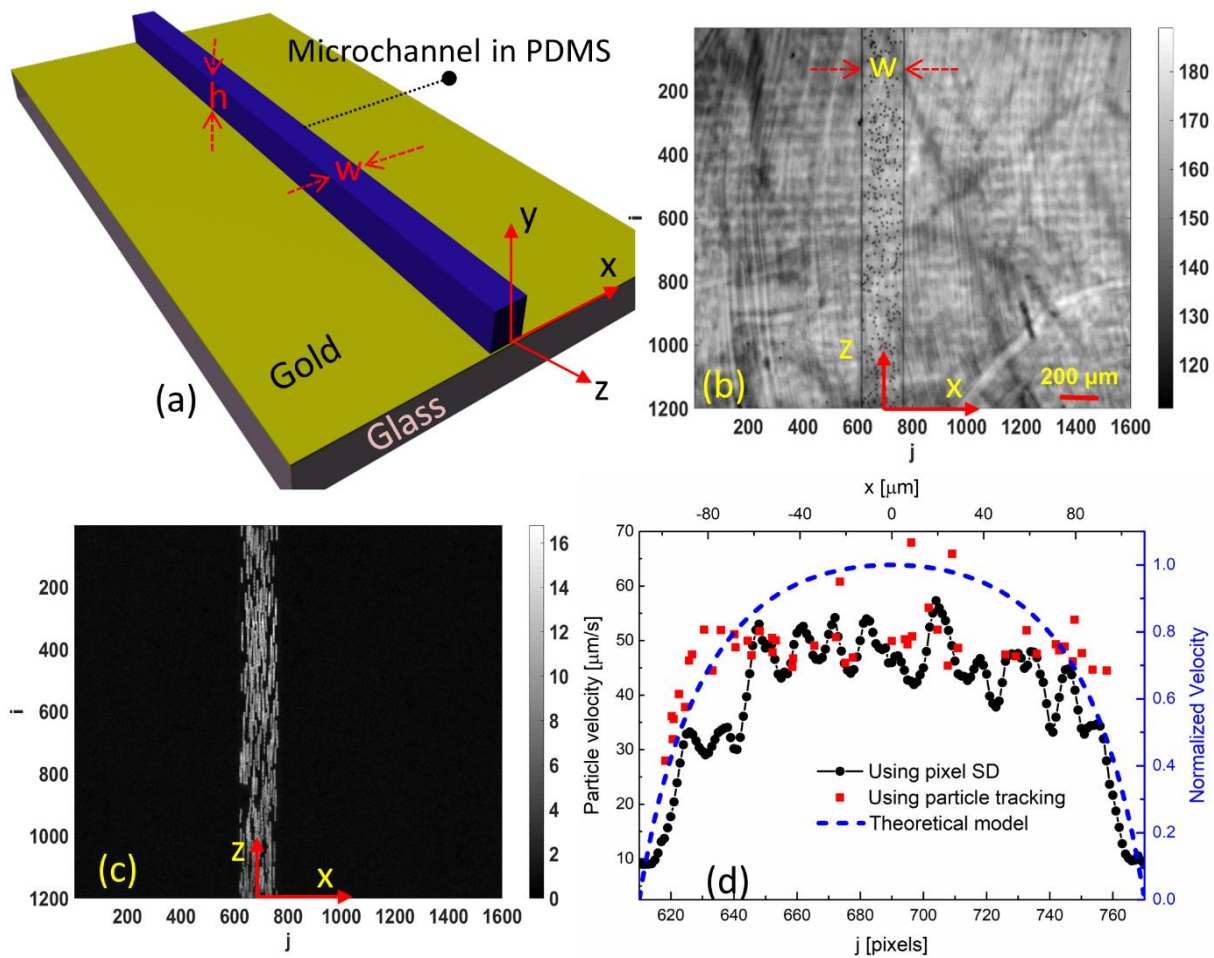

FIG. 3. (a) Schematic of the single microchannel device. (b) Microscopic image of the fabricated device. (c) Standard deviations of the pixels for a time period of 1.5 seconds (15 frames). The imaging is done on an area of 1600 pixels wide and 1200 pixels long (d) Velocity profile along the x direction of the microfluidic channel. The black circular markers indicate the velocity estimated by from the pixel standard deviation (SD). The red square markers show the velocity estimated by particle tracking. Blue, broken line is the normalized velocity profile as predicted by the Fourier Sum approximations[19].

**Characterizing direction of particles**

For creating flows at varying directions, a second device with structure as shown in FIG. 4 (a) was made. It consisted of two microfluidic channels (h = w = 50 µm) as in the previous device. The micro channels were then merged into a wider region (W' = 1 mm) so that the flow direction changes as shown in video [SI video2]. Figure 4 (b) shows the microscope image of the second device. As mentioned in the previous section, the experiments consisted of observing, analyzing and extracting information from the images. Initially, a standard deviation plot was generated as in the previous case. For extracting the local direction of the particles, we subdivided the imaging area into a 25×25 array of equal sizes. In each subdivision, the prominent direction was found out by the directionality plugin of ImageJ[20] using the scale space approach to the directional analysis [21]. FIG. 4 (c) is the directionality plot generated by replacing each of the 25×25 elements of the array with quivers at the angles extracted from the analysis and with standard deviation plot superimposed on it. FIG. 4 (d) is the magnified image for the regions marked with broken lines in FIG. 4 (c). The change is the flow direction as the particles enter the wider region is apparent from this image.

FIG. 4 (e) is the average direction of particles at selected x positions. It can be seen that the particles move in z direction, at an angle closer to $\frac{\pi}{2}$ with respect to XY plane before reaching the wide opening (z=0 to z=470). Inside the wider opening (z>470), the angles spread out towards 0 showing an orthogonal component of the flow, in addition to the initial flow parallel to the z-axis.

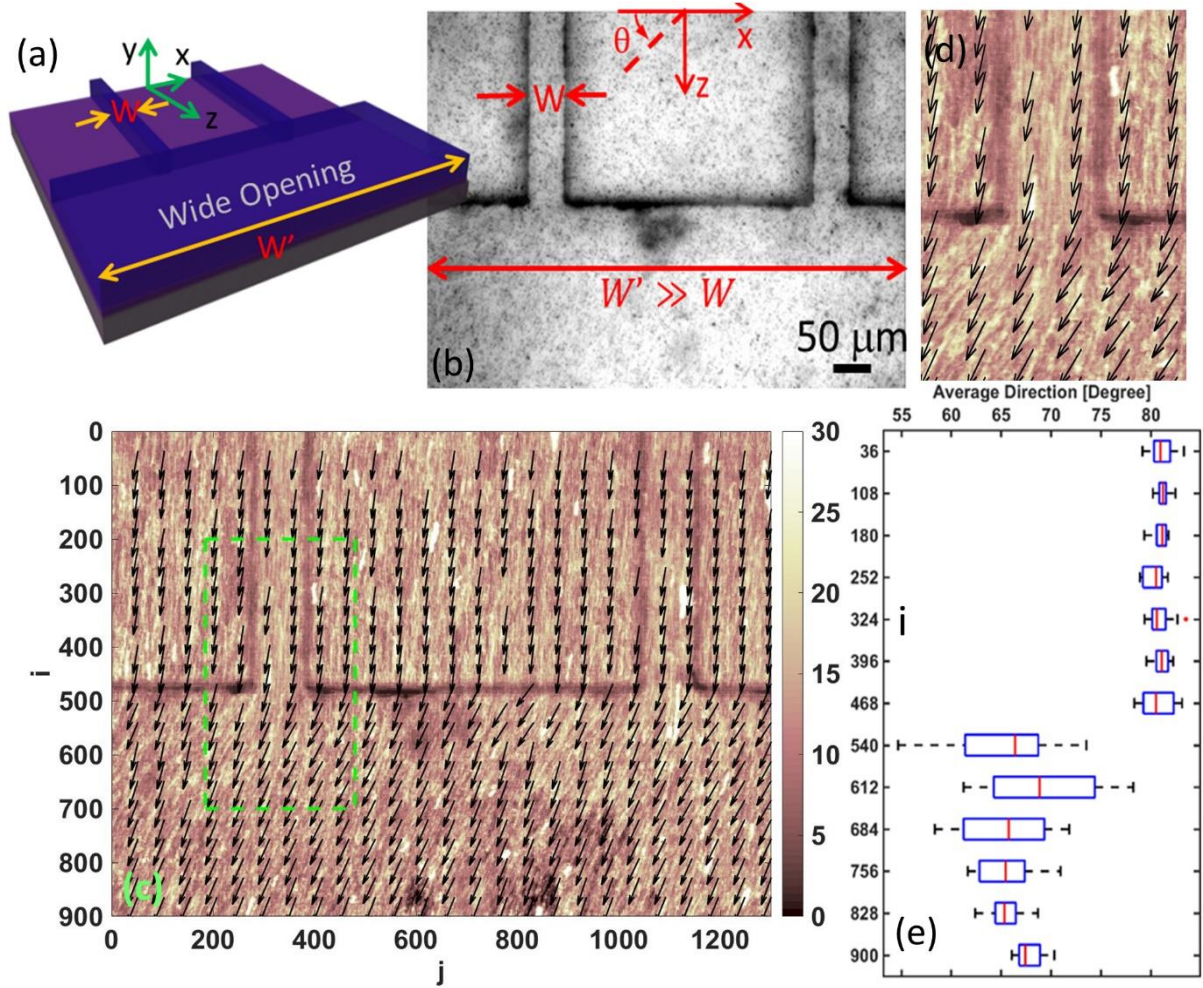

FIG. 4. (a) Schematic of the microchannel device opening to a wider area. (b) Microscopic image of the fabricated device. (c) Standard deviations of the pixels for a time period of 10 seconds (100 frames) superimposed by the quiver plot showing the direction. (d) Shows the magnified image for the regions marked with broken lines in (c) (e) Direction of particle movement as a function of Z and averaged along X direction.

### **Characterizing particle-surface interactions**

From the theoretical model, it is understood that the total standard deviation (S) in a region is proportional to the density and velocity of the mobile particles. These parameters are in turn dependent on the net interaction force between micro particle and surfaces in contact. Hence, it can be hypothesized that the pixel standard deviation has a higher value for regions on a smooth substrate with lesser interaction (like Silicon) than that on rougher surfaces, such as charged polyelectrolyte layers, with higher particle-substrate interaction. Thus, we can use the pixel standard deviation to study the variations in particle motion and thereby the surface interactions. The following experiment is an illustration of such a potential application.

FIG. 5 (a) shows the schematic of the device used for this study. It consists of a silicon substrate, partially coated with a thin (total thickness of about 10 nm), 5 bilayer system of polyelectrolyte (PE) consisting of Polyallylamine hydrochloride (PAH) and Poly acrylic acid (PAA). It is generally represented as $(PAH/PAA)_5$. A fluid chamber is then made from a thin sheet of poly vinyl chloride (thickness of about 100 μm) with a hole (of diameter 3mm) in the centre [22]. The hole is aligned to the PE-Silicon interface. Micro beads in water medium (as in the previous experiments) were introduced into the chamber using a micropipette. The entire fluid column remains water tight with a cover slip attached at the top. FIG. 5 (b) shows the microscope image of the PE-Silicon interface in the fabricated device.

From the initial observations it was apparent that there was larger number of mobile particles over the bare silicon region than the PE region. The comparison of pixel standard deviations in the two regions revealed more qualitative and quantitative understanding of the effects of the surface interactions. For studying the velocity distribution of mobile particles, 14 regions of size 40 pixel × 40 pixel were chosen, 7 each on PE and silicon. Each of these regions contains the entire trajectory of one of the mobile particles for a time duration of 30 seconds (300 timeframes). The histogram of the pixel standard deviations was then calculated for each region. Average of seven of these histograms, were calculated and plotted in FIG. 5 (c) (details are given in SI text, Section 5). It can be seen that an average pixel on the PE region has a lesser standard deviation as well as a lesser span in standard deviation (and correspondingly, particle velocity). This shows that the mobile microparticles on Si move faster on average than those on PE.

Another interesting observation is obtained by comparing the time evolution of total standard deviation of pixel on PE and on bare Si. It has to be remembered that the sum of the pixel standard deviations in a region is proportional to the velocity and number of mobile particles in that region. For comparing the time evolution of total standard deviation of pixel on PE and on bare Si, regions of dimensions 1200 pixel high and 600 pixel wide were chosen on either surfaces (Refer FIG. 5 (b)). To avoid the irregularities at the interface, the remaining central region (1200 pixel high and 400 pixel wide) was neglected. The total standard deviation for a time duration of 30 seconds were calculated in every 2 minutes. Portions from these observations are given in SI video3. FIG. 5 (d) shows the evolution of total pixel standard deviation ($S$) at regions on either surface. Fitting the evolution of total standard deviations to exponential decay models, a faster decay time constant (13.3 minutes) was extracted for PE than that for Silicon (22.2 minutes) surface. This points towards a higher damping of mobile particles on rough PE than those on Silicon as expected from comparison of surface roughness between these two structures shown in SI section 6. It is seen that the PE surface is significantly rougher indicating a larger surface interaction with the particles and consequently greater damping of its motion as revealed by the temporal evolution of the pixel variance.

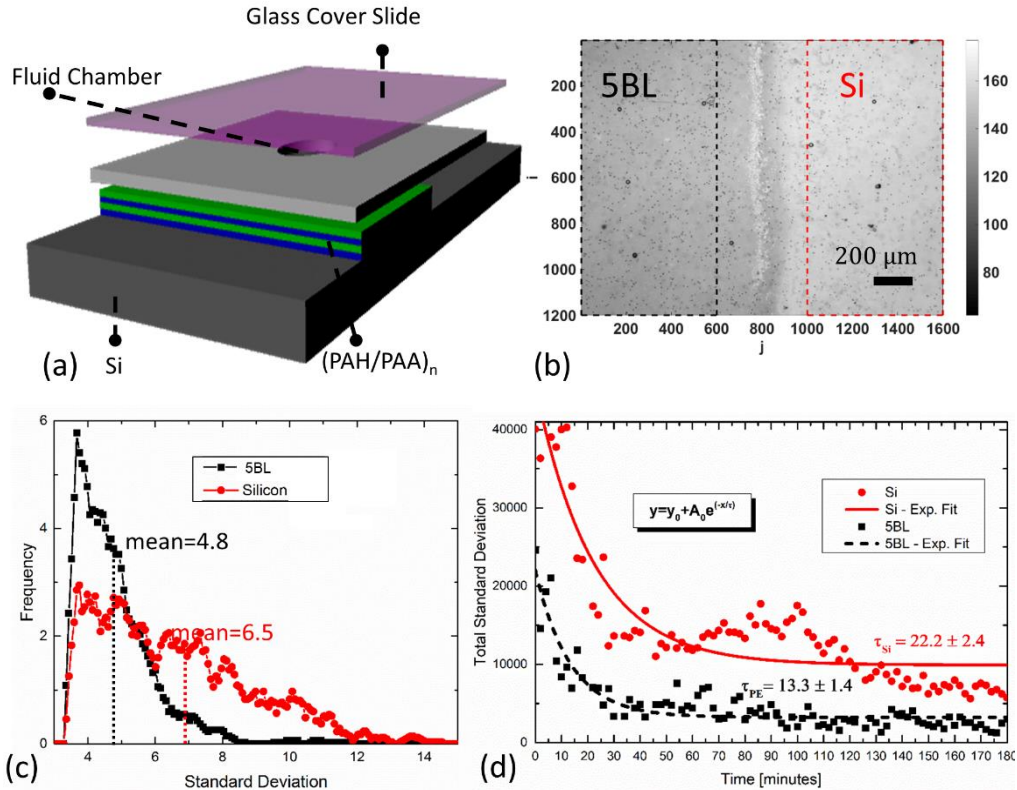

FIG. 5. (a) schematic of the device used to compare microparticle-surface interacions on Polyelectrolyte (PE) multilayer and on bare silicon. (b) Microscope image of the PE multilayer – bare silicon interface. (c) histogram of pixel standard deviations on an average 40 pixel ✕ 40 pixel region of particle trajectories on PE and on Si. (d) The time evolution of total standard deviation (for 300 frames) of pixels in regions of 1200 pixel ✕ 600 pixels on PE multilayer and on Si.

## Conclusions

To summarize, we have developed a simple, tracking-free method using variations in image space to analyse the speed, direction and surface-interaction of micro particles in various microfluidic systems. Velocity profiles closely matching theoretical estimates were obtained using this technique. This technique can be used in calculating flow patterns without any additions to the standard microscope setup. This could be a novel method in analysing the movements of microorganisms or microparticles without resorting to single particle tracking. Further, we also demonstrated the potential use of this method in probing microparticle-surface interactions which can be explored in future for sensing applications. We intend to explore the application of this technique to probe bacterial motility, especially to extract local structures within bacterial flows such as localized vortices. It is hoped that the image variance based method we described here presents an addition to the suite of tracking-free techniques such as Differential Dynamic Microscopy (DDM) to extract motility parameters from video data.

# Supplementary Information
# Tracking-free Determination of Microparticle Motion from Image Variance

Harish Sasikumar[1], and Manoj M. Varma[1,2,1]

[1]Center for Nano Science and Engineering, Indian Institute of Science, Bangalore, 560012, India

[2]Robert Bosch Center for Cyber Physical Systems, Indian Institute of Science, Bangalore, 560012, India

## 1 Useful Formulae

$$\int_{-\infty}^{\infty} e^{-\frac{x^2 - Ax}{R^2}} dx = R\sqrt{\pi}\, e^{\frac{A^2}{4R^2}} \quad (1)$$

$$\int_{-\infty}^{\infty} e^{-\frac{x^2}{R^2}} dx = R\sqrt{\pi} \quad (2)$$

## 2 Image of a micro particle: Comparing Experimental observation and Simulation model

### 2.1 Theory

Though the point spread function of a diffraction-limited case is an Airy distribution, for mathematical conveniences, it can be approximated to a Gaussian distribution[1].

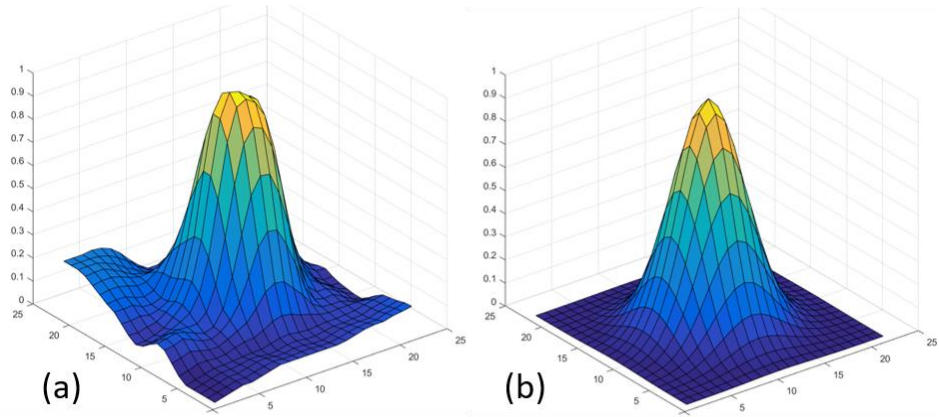

**Figure 1 Comparing Gaussian Plot and particle profile**

(a) Typical 10-micron polystyrene bead imaged with a 10x objective. The observation area is $22 \times 22$ pixels and the profile is smoothened, inverted and normalized. Using curve fitting, $\sigma$ was found to be ~3.2 (b) Normalized Gaussian plot with $\sigma = 3.2$ and center at (11.5,11.5)

For smoothening the image of micro particle, before fitting to Gaussian, the code given in [2] is used with a window size of 3 pixels.

---

[1] Author to whom correspondence should be addressed. Electronic mail: mvarma@iisc.ac.in

# 3 Linear motion of multi-pixel particle with Gaussian profile

## 3.1 Theory

Consider the image of a particle with centre at $(i_r, j_r)$ and standard deviation $R$, creating an intensity distribution in the image plane as follows

$$x_{i,j}^{(G)} = Ae^{-\frac{(i-i_r)^2+(j-j_r)^2}{2R^2}} \quad (3)$$

As shown in Figure 2(a), the particle moves from its initial position to the final position (along j axis) in $N$ time frames. The centre of the particle transverses m pixels during this time. Figure 2(b) shows the corresponding standard deviation obtained in the image plane. (Note that irrespective of the representation of the particle – as dark entity in a bright field or bright entity in a dark field- the standard deviation will be the same)

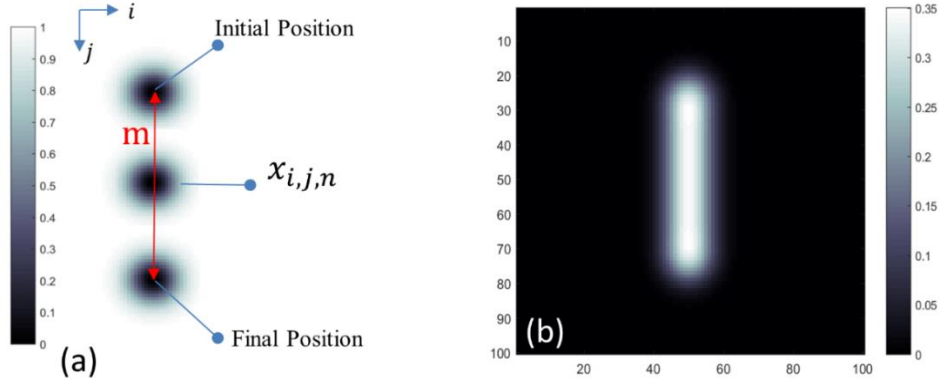

**Figure 2 Multi-pixel particle performing linear motion**

(a) Particle at its initial, final and an in-between position (A=1). (b) a typical standard deviation plot of the linear motion.

The value of a pixel at the $i^{th}$ position along the centre of the particle (at time $n$) is given by

$$x_{i,j,n} = Ae^{-\frac{i^2}{2R^2}} \quad (4)$$

At any other time-instance, k frames ahead, it can be given as

$$x_{i,j,n+k} = Ae^{-\frac{(i-k)^2+k^2}{2R^2}} \quad (5)$$

Hence the average value of a pixel, away from the initial or final positions can be approximated as,

$$\overline{x_{I,J}^{(G)}} = \frac{A}{N}\sum_{k=-\infty}^{\infty} e^{-\frac{(i-k)^2+k^2}{2R^2}} = \frac{A}{N} e^{-\frac{i^2}{2R^2}} \sum_{k=-\infty}^{\infty} e^{-\frac{k^2-ik}{R^2}} \quad (6)$$

Note that, even when the sum is composed of infinite number of terms, the dying nature of the function makes the sum finite. Approximating the sum as integral, and using equation 1, the above equation can be simplified into

$$\overline{x_{I,J}^{(G)}} = \sqrt{\pi}\frac{AR}{N} e^{-\frac{i^2}{4R^2}} \quad (7)$$

Similarly, the average of the squares of the pixel value can be calculated as

$$\overline{x_{i,j}^2} = \frac{A^2}{N} \sum_{k=-\infty}^{\infty} e^{-\frac{(i-k)^2+k^2}{R^2}} = \frac{A^2}{N} e^{-\frac{i^2}{R^2}} \sum_{k=-\infty}^{\infty} e^{-2\frac{k^2-ik}{R^2}} \tag{8}$$

Which can be simplified (by replacing $R$ with $R/\sqrt{2}$ in equation 7), to,

$$\overline{x_{i,j}^2} = \sqrt{\frac{\pi}{2}} \frac{A^2 R}{N} e^{-\frac{i^2}{2R^2}} \tag{9}$$

Thus, the variance of the pixel at position $(i,j)$ can be derived using equations 7 and 9 as,

$$\sigma_{i,j}^2 = \overline{x_{i,j}^2} - \left(\overline{x_{i,j}}\right)^2 = A^2 \left( \sqrt{\frac{\pi}{2}} \frac{R}{N} - \pi \frac{R^2}{N^2} \right) e^{-\frac{i^2}{2R^2}} \tag{10}$$

And for a particle which has moved m pixels, the total standard deviation is given (using the approximation in equation 2) by

$$\sigma = m \sum_{i=-\infty}^{\infty} \sigma_{i,j} = mA \sqrt{\left( \sqrt{\frac{\pi}{2}} \frac{R}{N} - \pi \frac{R^2}{N^2} \right) R\sqrt{\pi}} \tag{11}$$

Note that, $m$ arises due to the length of the variance plot and the summation, due to the width of the particle. For $p$ particles with non-intersecting trajectories, the total standard deviation is given by,

$$S = p\sigma = Apm \sqrt{\sqrt{\frac{\pi^3}{2}} \frac{R^3}{N} \left(1 - \sqrt{2\pi} \frac{R}{N}\right)} \tag{12}$$

$m$, the length of the variance plot arises due to the movement of particle and hence is related to the apparent velocity ($v_s$ [pixel/frame]) of the particle and the total time frames $N$, by the following relation

$$m = Nv_s$$

The apparent velocity of the particle in the image space is related to the actual velocity ($v$) by the scaling factors of the video imaging system: the dimension scaling factor $\gamma \left[\frac{\mu m}{pixel}\right]$ and the frame acquisition rate $f \left[\frac{frames}{second}\right]$.

$$v_s = \frac{v}{\gamma f}$$

Hence, $S$ can be written as,

$$S = \frac{Apv}{\gamma f} \sqrt{\sqrt{\frac{\pi^3}{2}} NR^3 \left(1 - \sqrt{2\pi} \frac{R}{N}\right)} \tag{13}$$

### 3.2 Simulation

Using equation 13, the velocity of the particles in a rectangular channel can be calculated from the total standard deviation ($S$) calculated by summing in the vertical direction.

$$v = \frac{\gamma f S}{ApC}; \quad C = \sqrt{\sqrt{\frac{\pi^3}{2}} NR^3 \left(1 - \sqrt{2\pi}\frac{R}{N}\right)}$$

*Values used for the simulation*

| Parameter | Definition | Value or estimate |
|---|---|---|
| $\gamma$ | dimension scaling factor | 1.39 $\mu m/pixel$ (Microscope specification) |
| $f$ | Frame rate of image acquisition | 10 frames/ second |
| $A$ | Height of the Gaussian (image of the particle) | 33.58 (average obtained from the images of 5 particles) |
| $R$ | Standard deviation of Gaussian (image of the particle) | 3.5 (average obtained from the images of 5 particles) |
| $p$ | Number of particles per unit pixel (along the width) | Total particles / width of the channel in pixels = 289 / 152 = 1.90 |
| $N$ | Number of time frames | 15 (1.5 seconds of data at 10 frames per second) |

# 4  Soft-Lithography for the fabrication of microfluidic devices

The processes for soft-lithography were adapted from [3]. The steps were as following

- PDMS elastomer and curing agent (SYLGARD 184, from Dow Corning) were taken in the weight ratio 10:1 in a clean, disposable cup. They were mixed thoroughly using a spatula for ~5 minutes and the mixture is kept in a low-pressure chamber (desiccator connected to a vacuum pump) for ~1 hour to degas out the bubbles.
- The degassed mixture was poured onto the mold (master fabricated on silicon substrate) kept in a Teflon Petri dish and once again degassed. It was then cured at 90°C for ~3 hours over a hot plate. After curing, the PDMS structures were allowed to cool down to room temperature.
- The PDMS structures were then peeled-off carefully from the mold. A biopsy punch of was used to drill holes at the outlet and inlet points.
- The freshly prepared PDMS surfaces were exposed to Oxygen plasma for ~30 seconds using a 150 W RF plasma generator and with atmospheric air. The exposed PDMS surface were kept onto a gold coated glass slide and a small pressure was applied using a plastic tweezer. Further the bonded samples are baked at ~80°C for 20 minutes.

# 5  Histograms of Static Particle and Dynamic Particles on Si and PE

The histograms were thresholded for pixel values above 3.5 to reject the inherent noises in the imaging system, including the CMOS sensors. Refer Figure 3 where the standard deviation of a 40 pixel × 40 pixel region involving a static microparticle is given.

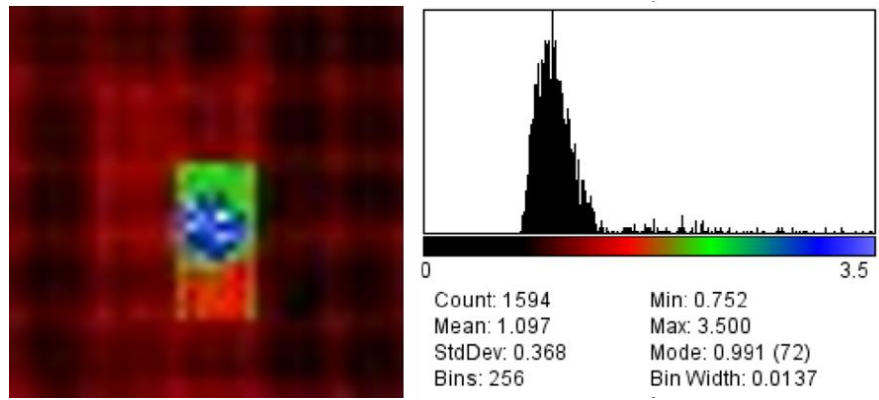

**Figure 3. Standard deviation of a static particle**

Standard deviation (resulting from the noises in the imaging system) as of a 40 pixel × 40 pixel region involving a static microparticle.

14 regions, each containing the trajectory of single mobile particle for a time duration of 30 seconds (300 timeframes) were chosen. They were of size 40 pixel × 40 pixel and were chosen such that 7 were on Si and 7 were on silicon. The histogram of the pixel standard deviations were then calculated for each region. Average of seven of these histograms, were calculated and plotted. Figure 4 (a) and (b) show the histograms of particles on Si and PE regions respectively.

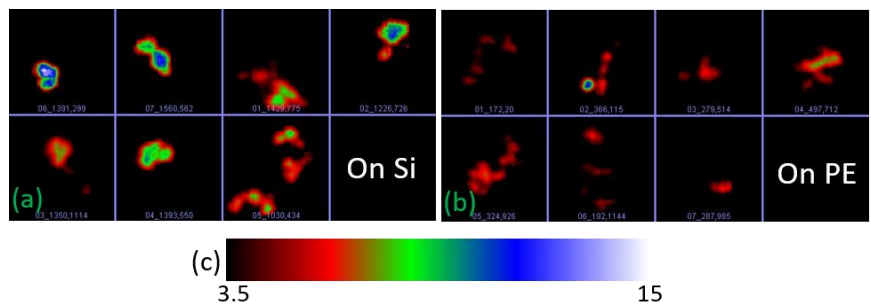

**Figure 4 Pixel standard deviation produced by mobile particles**

14 regions of size 40 pixel × 40 pixel were chosen, 7 each on PE and silicon. Each of these regions contains the entire trajectory of one of the mobile particles for a time duration of 30 seconds (300 timeframes). (a) and (b) are the pixel standard deviations of these regions in Si and PE, respectively. (c) is the colorbar of the standard deviation.

## 6 Comparison of Surface Roughness – Si v/s Polyelectrolyte

For the comparison of surface roughness, surface profile of patterned PEM (coated on Si sample) was obtained using an Atomic Force Microscope (AFM). A typical interface of the circular PEM pattern and Si is given in Figure 5.

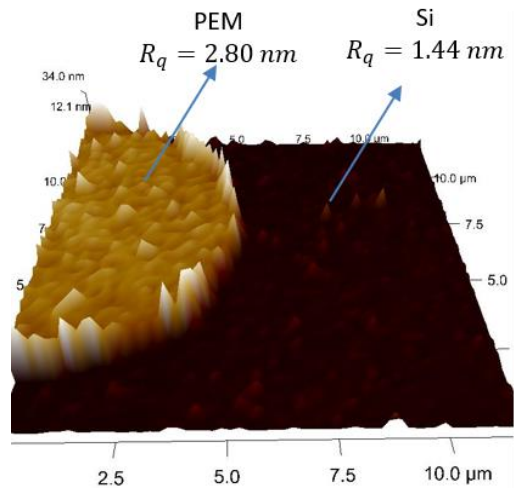

**Figure 5. Comparison of roughness in PE and Si regions**

Uniform regions of size $5\ \mu m \times 5\ \mu m$ were taken on the PEM and Si surfaces and the root mean square average of height deviation ($R_q$) was calculated. $R_q$ on PE and Si regions were found to be 2.80 nm and 1.44 nm respectively.